\documentclass[12pt,preprint]{aastex}
\begin{document}
\newcommand{\up}[1]{\ifmmode^{\rm #1}\else$^{\rm #1}$\fi}
\newcommand{\zdot}{\makebox[0pt][l]{.}}
\newcommand{\upd}{\up{d}}
\newcommand{\uph}{\up{h}}
\newcommand{\upm}{\up{m}}
\newcommand{\ups}{\up{s}}
\newcommand{\arcd}{\ifmmode^{\circ}\else$^{\circ}$\fi}
\newcommand{\arcm}{\ifmmode{'}\else$'$\fi}
\newcommand{\arcs}{\ifmmode{''}\else$''$\fi}

\title{The Araucaria Project. On the  Tip of the Red Giant Branch distance determination to the Magellanic Clouds. }

\author{Marek G{\'o}rski}
\affil{Millennium Astrophysical Institute, Santiago, Chile}
\affil{Universidad de Concepci{\'o}n, Departamento de Astronomia,
Casilla 160-C, Concepci{\'o}n, Chile}
\affil{Warsaw University Observatory, Al. Ujazdowskie 4, 00-478, Warsaw,
Poland}
\authoremail{mgorski@astrouw.edu.pl}

\author{Grzegorz Pietrzy{\'n}ski}
\affil{Nicolaus Copernicus Astronomical Center, Polish Academy of Sciences, Bartycka 18, 00-716, Warsaw, Poland}
\affil{Universidad de Concepci{\'o}n, Departamento de Astronomia,
Casilla 160-C, Concepci{\'o}n, Chile}
\authoremail{pietrzyn@camk.edu.pl}

\author{Wolfgang Gieren}
\affil{Universidad de Concepci{\'o}n, Departamento de Astronomia,
Casilla 160-C, Concepci{\'o}n, Chile}
\affil{Millennium Astrophysical Institute, Santiago, Chile}
\authoremail{wgieren@astro-udec.cl}

\author{M{\'a}rcio Catelan}   
\affil{Millennium Astrophysical Institute, Santiago, Chile}
\affil{Instituto de Astrof{\'i}sica, Facultad de F{\'i}sica, Pontificia Universidad Cat{\'o}lica de Chile, Av. Vicu{\~n}a MacKenna 4860, Santiago, Chile }
\authoremail{mcatelan@astro.puc.cl}

\author{Bogumi\l{} Pilecki}   
\affil{Nicolaus Copernicus Astronomical Center, Polish Academy of Sciences, Bartycka 18, 00-716, Warsaw, Poland}
\affil{Universidad de Concepci{\'o}n, Departamento de Astronomia,
Casilla 160-C,Concepci{\'o}n, Chile}
\authoremail{pilecki@astrouw.edu.pl}

\author{Paulina Karczmarek}
\affil{Warsaw University Observatory, Al. Ujazdowskie 4, 00-478, Warsaw,  
Poland}
\authoremail{pkarczma@astrouw.edu.pl}

\author{Ksenia Suchomska}  
\affil{Warsaw University Observatory, Al. Ujazdowskie 4, 00-478, Warsaw,
Poland}
\authoremail{ksenia@astrouw.edu.pl}

\author{Dariusz Graczyk}   
\affil{Nicolaus Copernicus Astronomical Center, Polish Academy of Sciences, Bartycka 18, 00-716, Warsaw, Poland}
\affil{Millennium Astrophysical Institute, Santiago, Chile}
\affil{Universidad de Concepci{\'o}n, Departamento de Astronomia,
Casilla 160-C, Concepci{\'o}n, Chile}
\authoremail{darek@astro-udec.cl}

\author{Piotr Konorski}
\affil{Warsaw University Observatory, Al. Ujazdowskie 4, 00-478, Warsaw,
Poland}
\authoremail{piokon@astrouw.edu.pl}

\author{Bart\l{}omiej Zgirski}
\affil{Nicolaus Copernicus Astronomical Center, Polish Academy of Sciences, Bartycka 18, 00-716, Warsaw, Poland}
\authoremail{bzgirski@camk.edu.pl}

\author{Piotr Wielg{\'o}rski}
\affil{Nicolaus Copernicus Astronomical Center, Polish Academy of Sciences, Bartycka 18, 00-716, Warsaw, Poland}
\authoremail{pwielgorski@astrouw.edu.pl}

\begin{abstract}

We present a precise optical and  near-infrared determination of the Tip of the Red
Giant Branch (TRGB) brightness in the Large and Small Magellanic Clouds (respectively LMC and SMC).
The commonly used calibrations of the absolute magnitude of the TRGB lead to an
overestimation of the distance to the LMC and SMC in the K band, and an underestimation
of the distance in the optical I band for both galaxies. Reported discrepancies are
at the level of 0.2 mag, with respect to the very accurate distance determinations to both Clouds based
on late-type eclipsing binaries. The differential distances between the LMC and SMC 
obtained in the J and K bands, and for the bolometric brightness are consistent with each other, 
and with the results obtained from eclipsing binaries and other distance indicators. 

\end{abstract}


\section{Introduction}
The main goal of our ongoing Araucaria Project is an accurate calibration of the extragalactic distance scale in the local universe. Since the first publications resulting from this project (Pietrzy{\'n}ski \& Gieren 2002, Pietrzy{\'n}ski et al. 2002) we have investigated numerous commonly used standard candles, including the Cepheid period-luminosity relation, the RR Lyrae period-luminosity-metallicity relation, the mean brightness of the red clump stars and the Tip of the Red Giant Branch (TRGB) magnitude in optical and near-infrared bands (Gieren et al. 2005; Szewczyk et al. 2008; Pietrzy{\'n}ski et al. 2009, 2010). Each of these methods of distance measurement have their own and unique potential of application, 
arising from different distance ranges, observational difficulties and sources of systematic error limiting the achievable accuracy. Measurements are disturbed by numerous effects, like reddening or differences of the chemical composition and age of stellar populations. Each of these factors affects distance measurements in a different way, depending on the technique that was used, as well as the band in which the observations were carried out. As a result, discrepancies between the distances obtained for the same galaxy from different methods can reach 20\%, as in the case of the M33 galaxy (Gieren et al. 2013). Identifying and correctly assessing the impact of population effects is an indispensable step in setting up an accurate distance scale of the universe.
 
In this paper we focus attention on the TRGB method. The concept of using the TRGB as a distance indicator was proposed by Baade in 1944, and later by Sandage in 1971. The first attempts to measure distances with this technique were made by Mould, Kristian \& Da Costa (1983, 1984) and Graham (1982). 
Subsequent studies showed that the absolute brightness of the TRGB in the optical I band depends very little on metallicity. This dependence was calibrated with extensive observations of red giants in Galactic globular clusters (Da Costa \& Armandroff 1990). In 1993 Lee, Freedman \& Madore investigated the accuracy of distance measurement methods, by comparing measurement results for 10 Local Group galaxies. They used three different standard candles: Cepheids, RR Lyrae stars and the TRGB. The general conclusion from this comparison was that all three methods yield results which are consistent within 0.1 mag. Recent studies confirm that distance measurements based on the TRGB do not depend on the shape of the star formation history (SFH), or the age of the stars if the mean metallicity of the red giant branch (RGB) is smaller than -0.3 dex and the stellar population is older than 1.5 Gyr (Kennicutt et al. 1998; Ferrarese et al. 2000; Barker et al. 2004). To date, the distances to more than 40 galaxies nearer than 10 Mpc have been obtained with the I band TRGB brightness (Karachentsev et al. 2003; Jacobs et al. 2009). 

However, like many other techniques, the TRGB measurements can be affected by reddening. This effect is identified as one of the most important systematic errors on distance measurements. 
Important progress was made in 2004 when Valenti, Ferraro \& Origlia calibrated the TRGB infrared J, H and K absolute brightnesses in terms of metallicity. Near-infrared photometry minimizes the influence of interstellar reddening. Unfortunately, however, the TRGB brightness in the near-infrared bands is very sensitive to metallicity (a 0.1 dex change in metallicity changes the brightness of the TRGB in the K band by 0.058 mag). At present, the accuracy of distance measurements based on the J and K band photometry of the TRGB is not sufficiently well studied, and distances to only a few galaxies have been obtained using infrared bands. Recently, our group extended the list of near-infrared based TRGB distance measurements to the Carina, Fornax, Sculptor Dwarf Spheroidal galaxies (Pietrzy{\'n}ski et al. 2009), and NGC 6822, NGC 3109, IC 1613, and WLM (G{\'o}rski, Pietrzy{\'n}ski, Gieren 2011). 

It is certain that the TRGB method has a great potential of application. The reliability of the method in the optical I band has been shown by many studies and the near-infrared measurements offer the opportunity to minimize the reddening problem. 
However, there is an open question about the limitation and uncertainty on the distances obtained from these measurements. The TRGB method was successfully used to measure the distances to galaxies with red giant branches dominated by a homogeneous and old stellar population. Salaris and Girardi (2005) pointed out that in case of composite stellar populations, distance measurements with the TRGB technique can be affected by systematic errors. By employing theoretical stellar population synthesis techniques, they studied galaxies with complex Star Formation Histories (SFHs) and Age-Metallicity Relationships (AMRs), namely the Large and Small Magellanic Clouds  (respectively LMC and SMC). They concluded that the TRGB distance measurements to Magellanic Cloud-type galaxies can be affected by up to 0.1 - 0.2 mag when using calibrations based on Galactic globular clusters.

The LMC and SMC are the two most extensively studied galaxies, and with more than 500 distance determinations from a variety of techniques (de Grijs et al.  2014) they are a natural choice for studies aimed at testing the accuracy of the TRGB method. The recent very accurate 2.2\% distance measurement to the LMC from eclipsing binaries (Pietrzy{\'n}ski et al. 2013) gives us a unique opportunity to compare different methods, and use the LMC as a very accurate zero-point for distance calibrations. Measurements of the TRGB brightness in the optical I band were performed by Reid et al. (1987), Sakai et al. (2000) and Cioni et al. (2000). The reported brightness determinations differ from each other at the level of 0.3 mag.

Studies of the near-infrared TRGB brightness in the LMC were performed by Kato et al. (2007) using the Infra-Red Survey Facility (IRSF), Nikolaev \& Weinberg (2000, using 2MASS), Cioni et al. (2000), and Macri et al. (2015). In the cases of Nikolaev \& Weinberg, Kato et al. and Macri et al., the mean brightness of the TRGB is reported without distance determination.
In the case of Cioni (DENIS survey), an advanced approach was adopted for the TRGB detection, and the distance determination is in excellent agreement with the 18.50 mag unreddened distance to the LMC coming from the late-type eclipsing binary studies of our group. However, these authors obtained the distance with a TRGB bolometric brightness calibration. Bolometric corrections in this case have been calculated from the K band and $(J-K)$ color with blackbody fits. The reported result are certainly valuable, but they cannot serve as the basis for the analysis of TRGB population effects.

Since the reported measurements are inconclusive in terms of statistical and systematic error assessment, we decided to thoroughly re-investigate the Large and Small Magellanic Clouds. Expected population effects should significantly differ in optical and infrared bands, which was the reason to focus on the optical I band, the near-infrared J, K bands, and bolometric brightness. Additionally, separate measurements of the TRGB magnitudes in 17 LMC fields make it possible to evaluate the TRGB intrinsic brightness variations for all analyzed bands. 

Our paper is organized as follows. The data sources and detection methodology
are described in the following section. Next, we present calibrations, and the resulting distance determinations to the LMC and SMC. Finally, we present a discussion of our results and a summary. 

\section{Photometric data} 

For our TRGB near-infrared analysis we used the IRSF Magellanic Clouds Point Source Catalog (Kato et al. 2007).  
The IRSF is a 1.4 m telescope, located at the South African Astronomical Observatory (SAAO) in Sutherland. The telescope is equipped with the SIRIUS camera (0.45 arcsec pixel scale). The photometric system consists of three near-infrared filters (J: 1.25 $\mu m$, H: 1.63 $\mu m$, K$_{\mbox{s}}$: 2.14 $\mu m$), similar to the 2MASS and UKIRT photometric systems. While there are other available surveys (e.g. DENIS and 2MASS) it is noticeable that IRSF provides a better angular resolution and deeper photometry than DENIS or 2MASS. 
For the optical analysis we used photometric maps of the OGLE-III survey (Udalski et al. 2008). Observations were carried out on the 1.3 m Warsaw telescope equipped with an eight-chip mosaic camera with 0.26 arcsec pixel scale. 
The Warsaw telescope is located at the Las Campanas Observatory, operated by the Carnegie Institution of Washington.  

\begin{figure}
\plotone{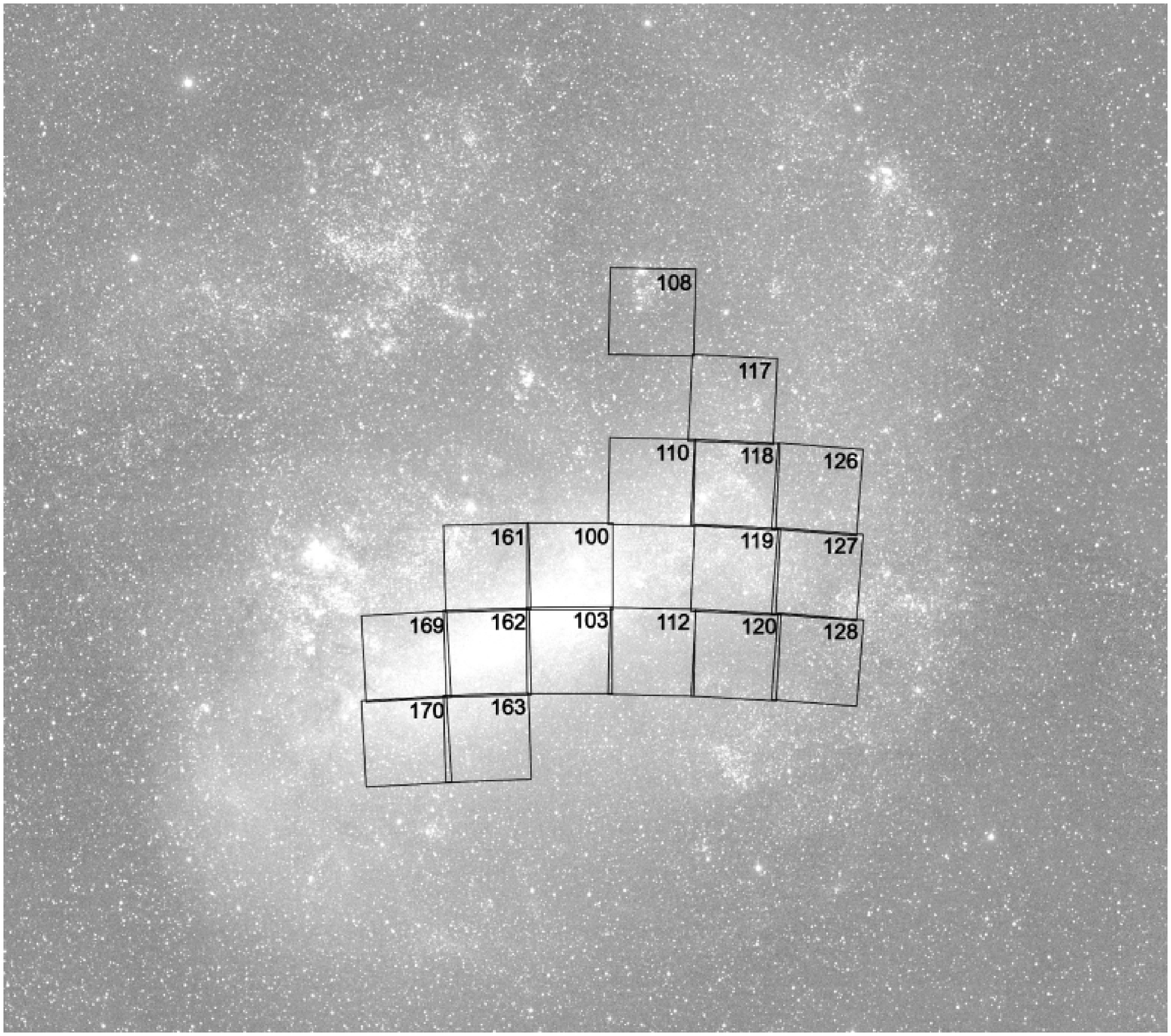}
\caption{The location of the 17 analyzed fields in the Large Magellanic Cloud. North is up, east is to the
left. Fields were selected to ensure a sufficiently large number of stars within 1 mag below the observed TRGB, and a minimal contribution to systematic errors from reddening and geometrical effects. The size of each field is 35 arcmin $\times$ 35 arcmin.}
\label{fig:lmc}
\end{figure}

In order to secure the most precise and reliable measurements possible, with minimal contributions from reddening and geometrical effects, we have selected fields in the central regions of the LMC and SMC, with internal reddening lower than E(V-I)=0.1 mag. The location of each of the 17 selected fields in the LMC and the five fields chosen in the SMC is shown in Figure \ref{fig:lmc} and Figure \ref{tab:smc}, respectively. The size of each field, both in the LMC and SMC, is  35 arcmin $\times$ 35 arcmin.

\section{TRGB detection}
Starting with the first publications on the TRGB method, many different techniques have been adopted in order to measure the brightness of the TRGB. 
In this paper we are using the Sobel filter, which was introduced by Lee et al. in 1993. 
It is based on the convolution of a luminosity function histogram with a discrete Sobel kernel [-2, -1, 0, 1, 2]. Subsequently this method was adopted for a continuous luminosity function (Sakai et al. 1996). A discrete histogram is replaced by a corresponding Gaussian-smoothed luminosity function $\Phi(m)$, which follows the expression: 

\begin{equation}
\Phi(m)=\sum_i^N \frac{1}{\sqrt{2\pi}\sigma_i}\exp\left[-\frac{(m_i-m)^2}{2\sigma^2_i}\right] \mathrm ,
\label{eq:sobel}
\end{equation}
where m$_i$ is the magnitude of the $i$-th star, $\sigma_i$ is the $i-th$ star photometric error, and $N$ is the total number of stars in the sample. Sobel filter answer $E(m)$ is defined as:
\begin{equation}
E(m) = \Phi (m+a) - \Phi (m-a) \mathrm ,
\end{equation}
where $a$ is the mean photometric error for all the stars with magnitudes between $m - 0.05$  and $m + 0.05$  mag. In this method, the brightness of the highest value of the Sobel filter answer is adopted as the observed brightness of the TRGB. 

\begin{figure}[htdp]
\epsscale{0.5}
\plotone{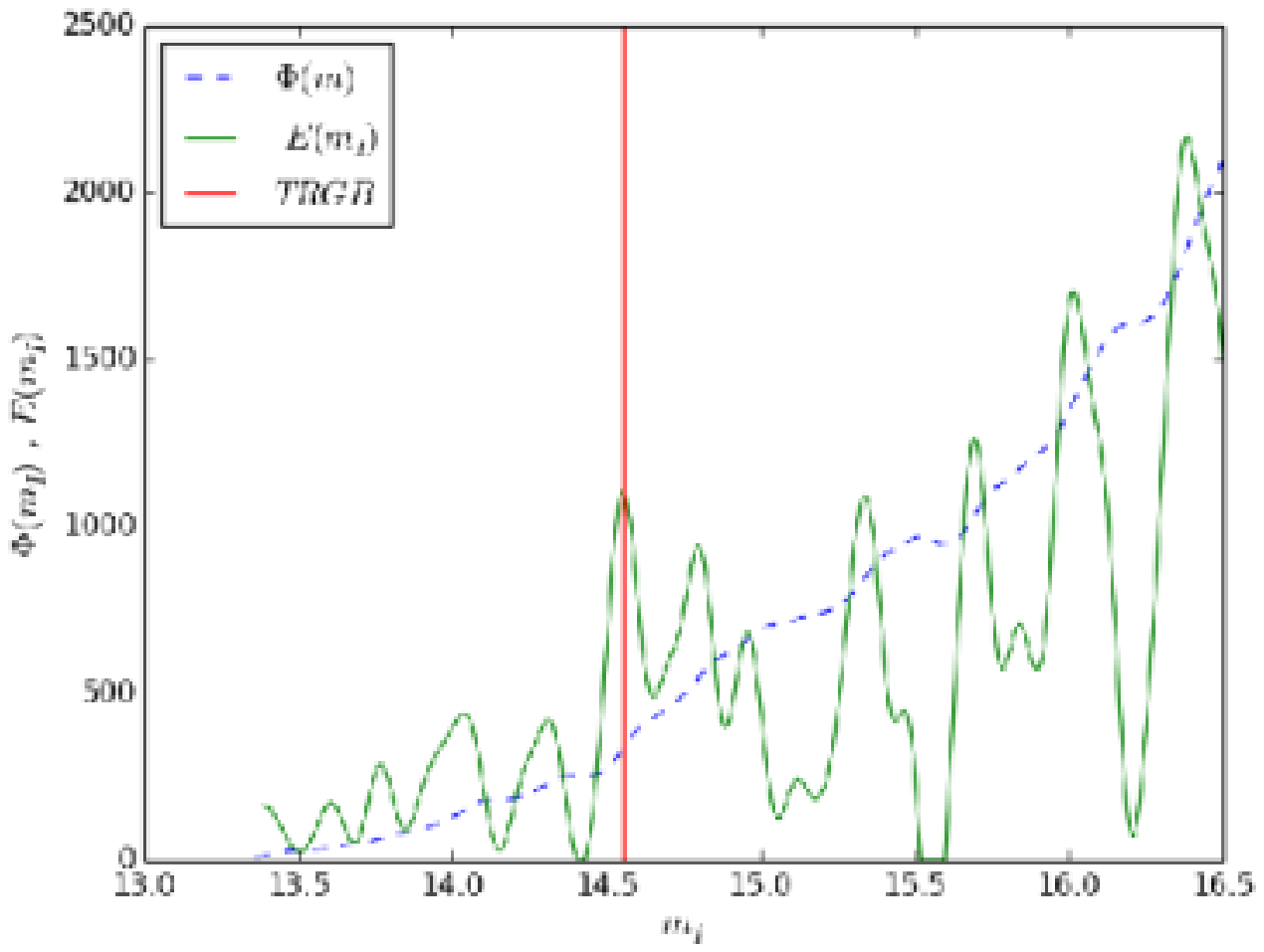}
\plotone{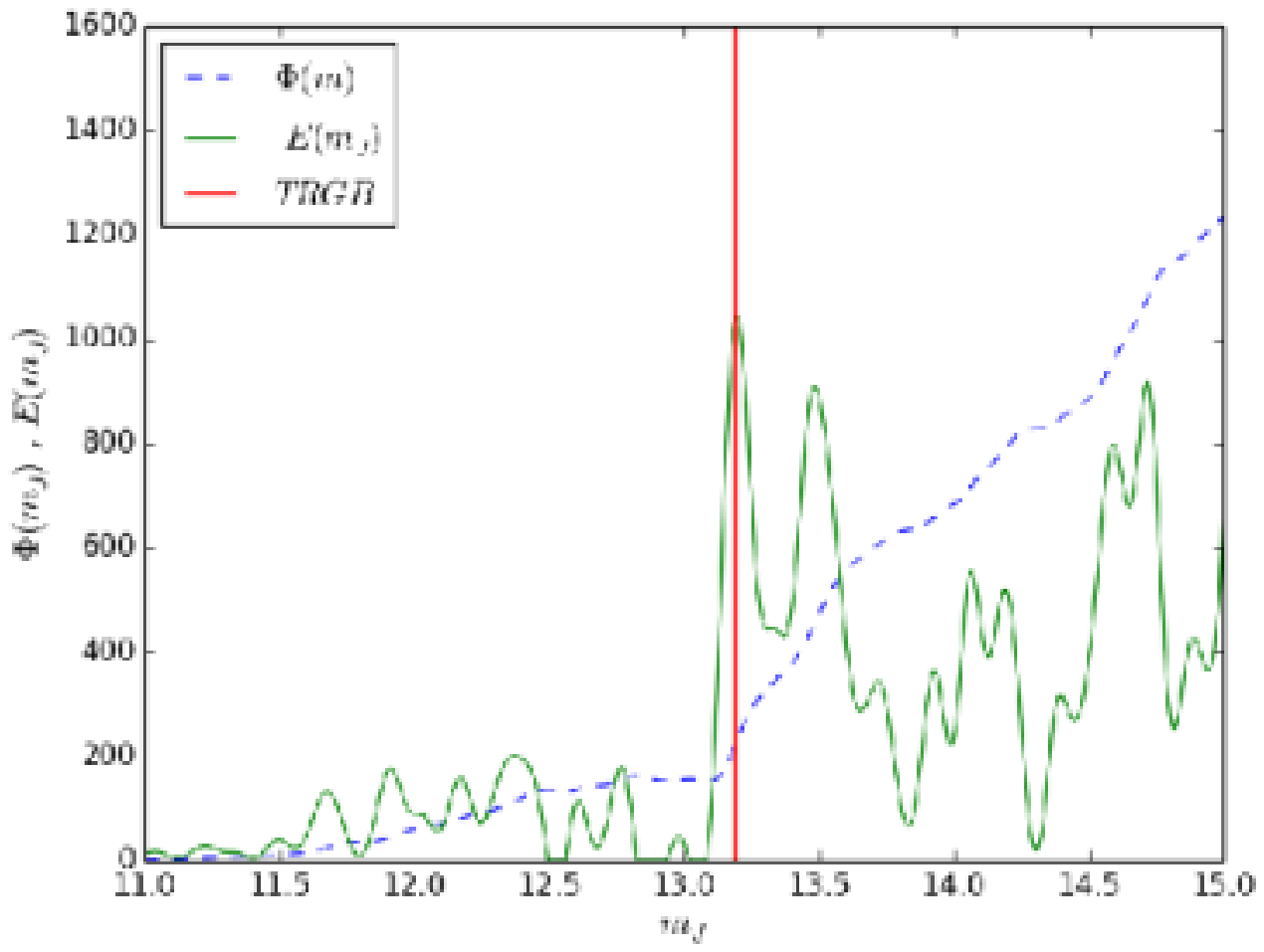}
\plotone{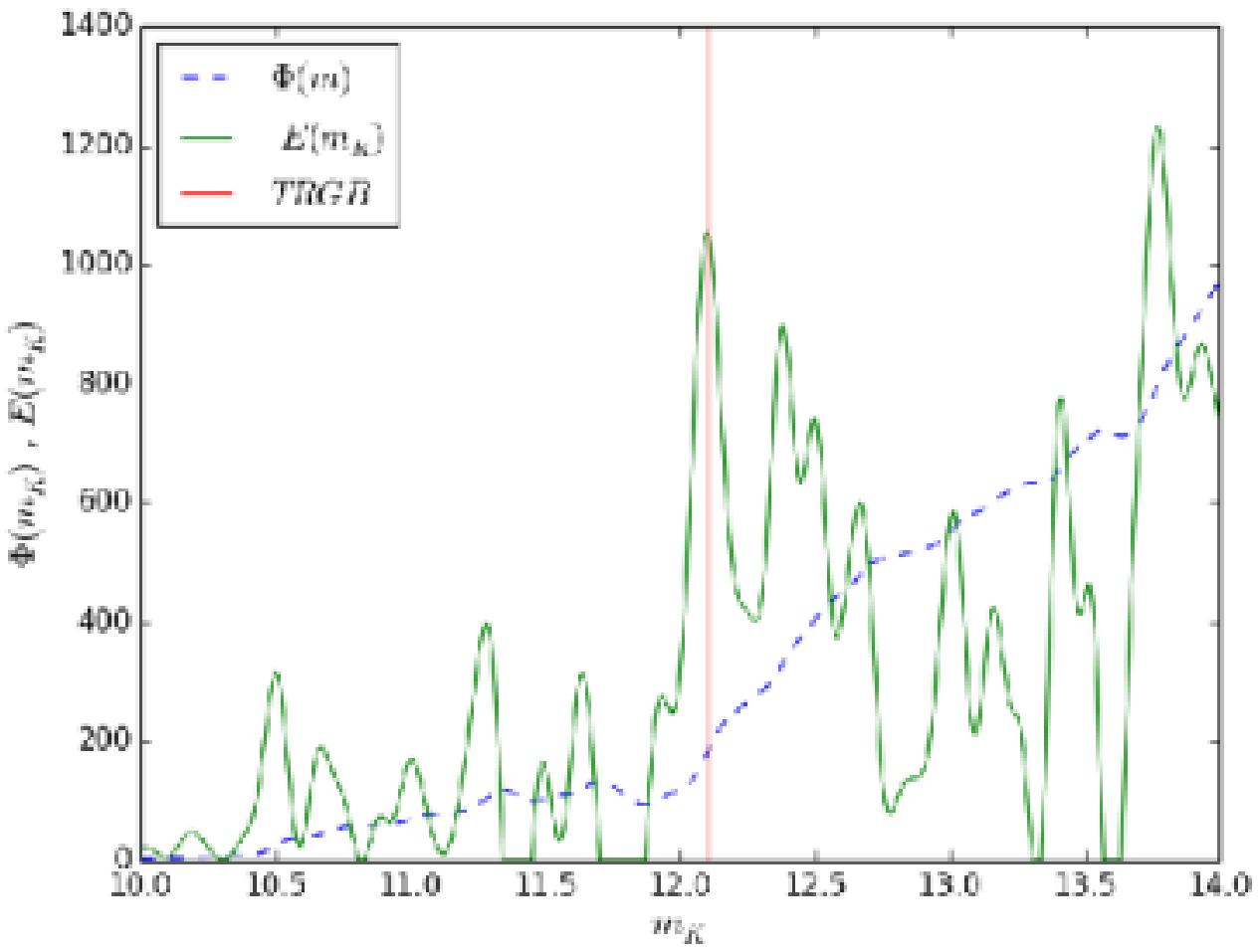}
\caption{The I, J and K band Gaussian-smoothed luminosity functions (blue dashed line) of the red giant branch in the Large Magellanic Cloud field lmc162, 
and the corresponding outputs of the Sobel edge-detection filter (green solid line). The vertical red line indicates the detected TRGB magnitude.}
\label{fig:lmc106}
\end{figure}

\begin{figure}[htdp]
\epsscale{0.5}
\plotone{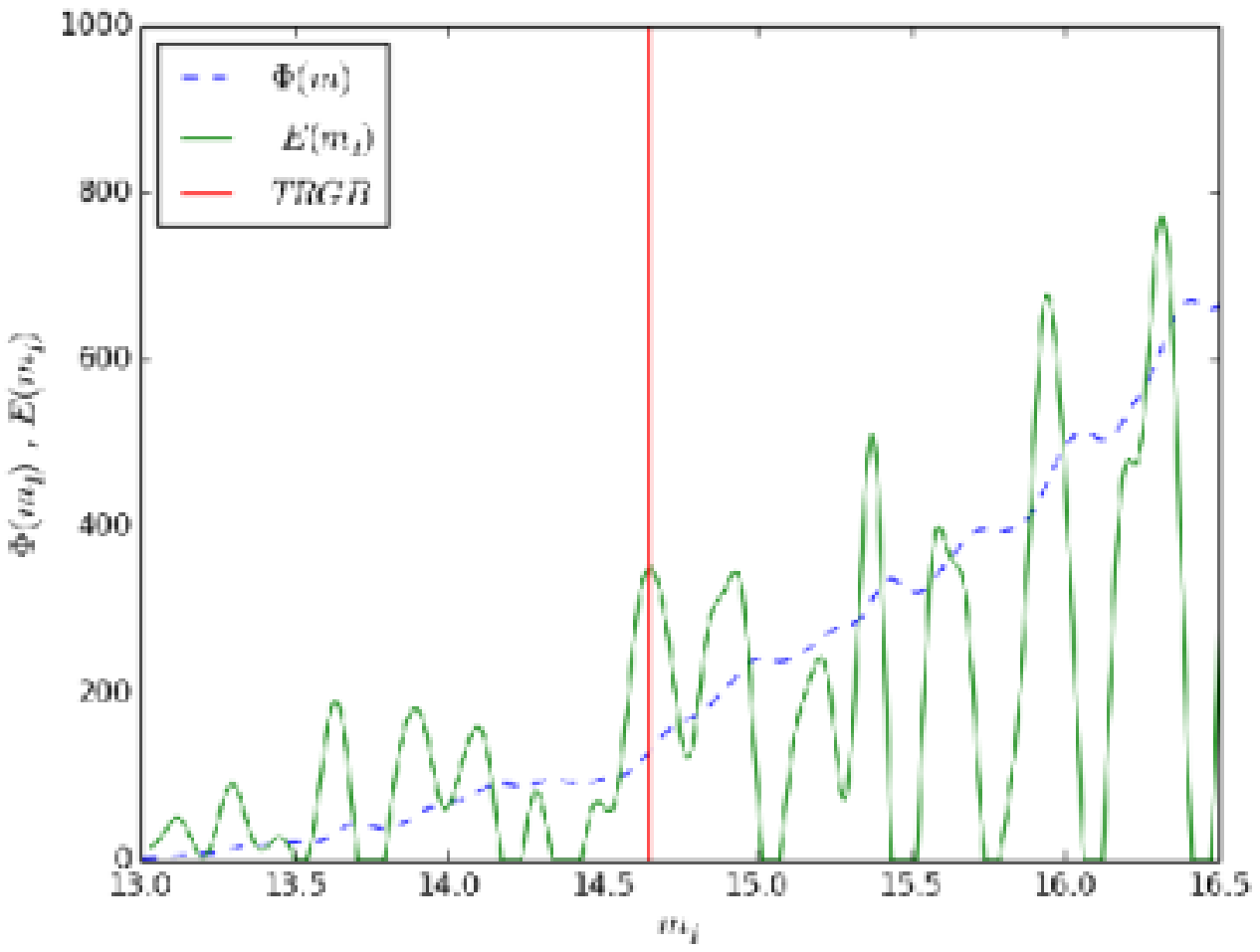}
\plotone{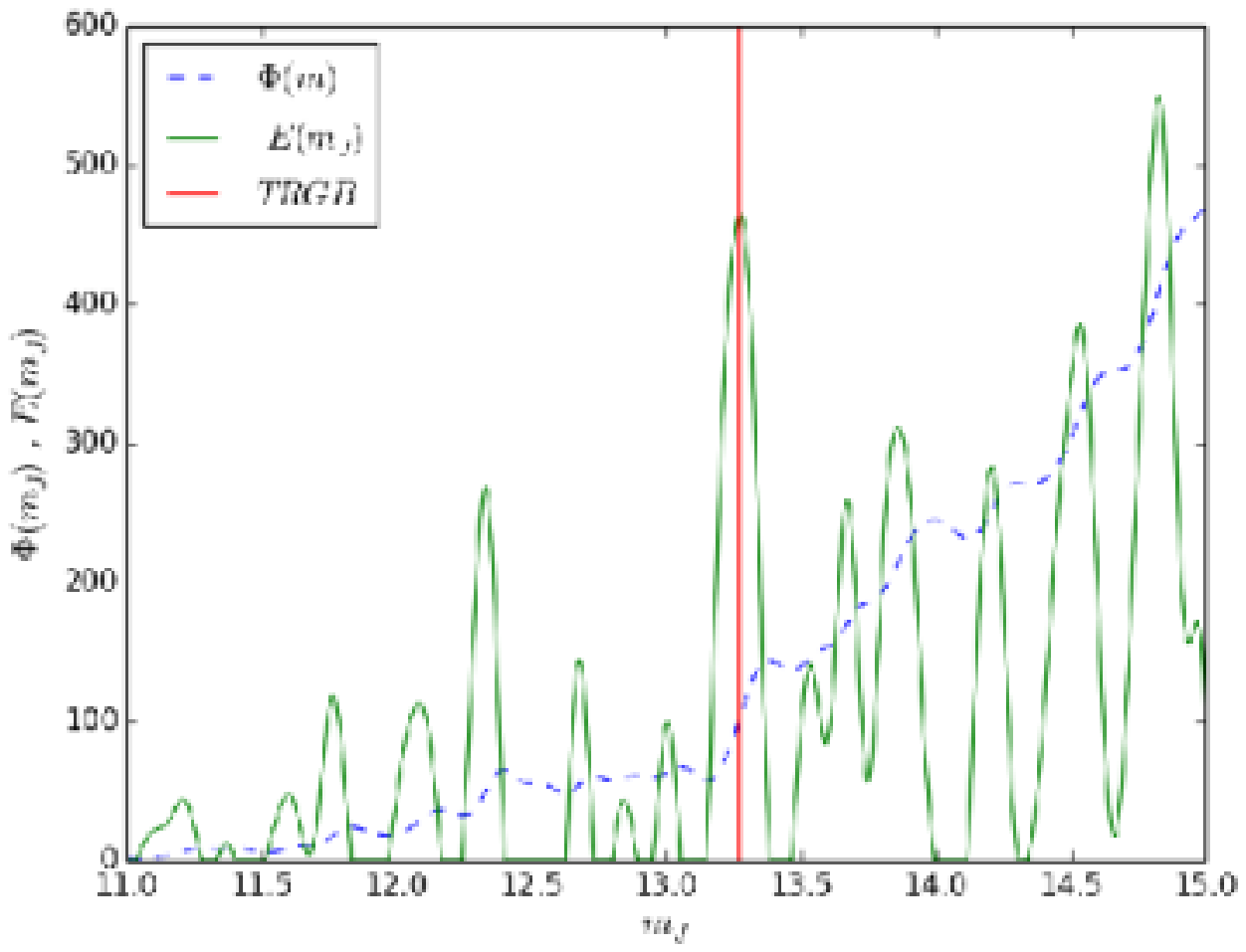}
\plotone{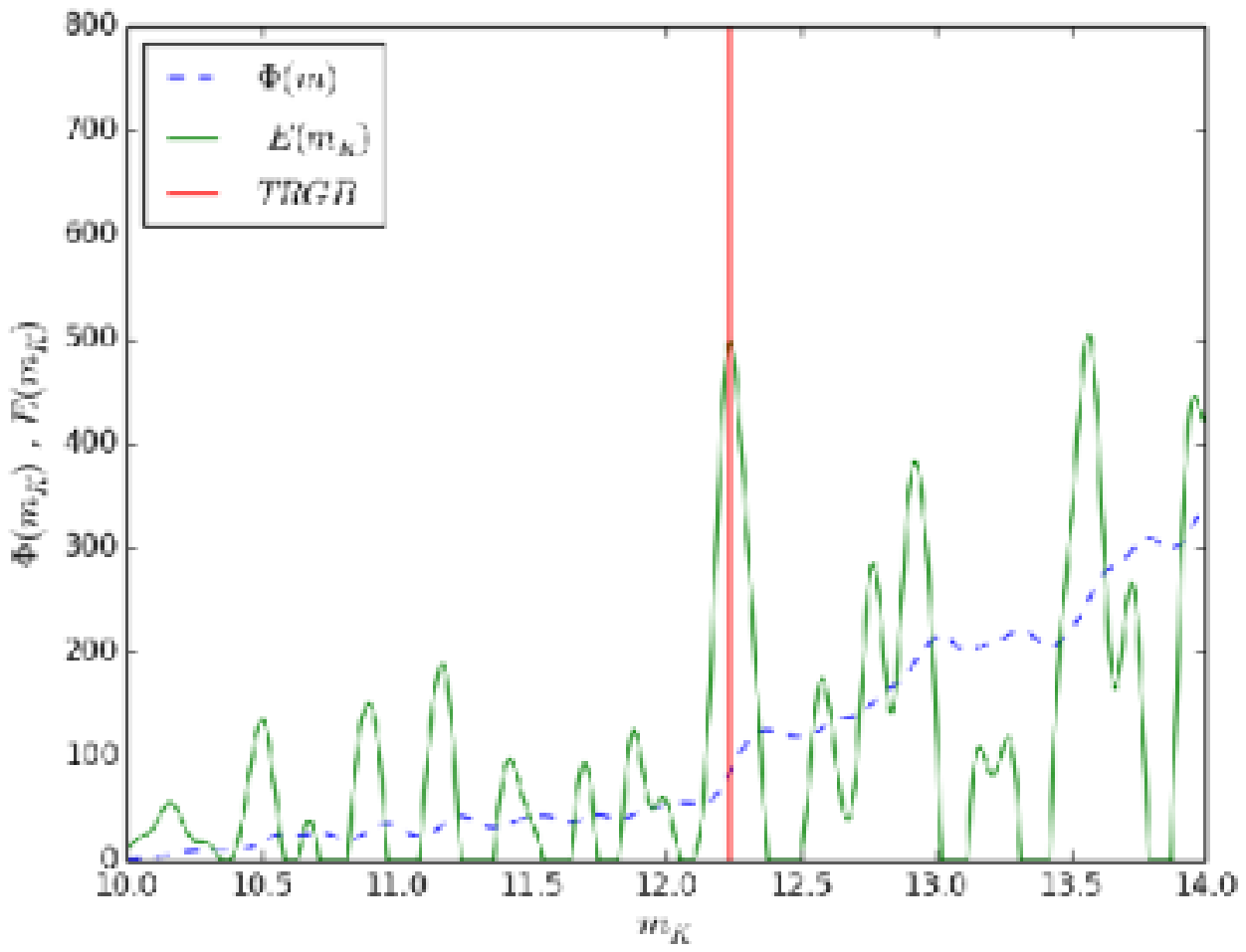}
\caption{The I, J and K band Gaussian-smoothed luminosity functions (blue dashed line) of the red giant branch in the Large Magellanic Cloud field lmc127, 
and the corresponding outputs of the Sobel edge-detection filter (green solid line). The vertical red line indicates the detected TRGB magnitude.}
\label{fig:lmc127}
\end{figure}

\begin{figure}[htdp]
\epsscale{0.5}
\plotone{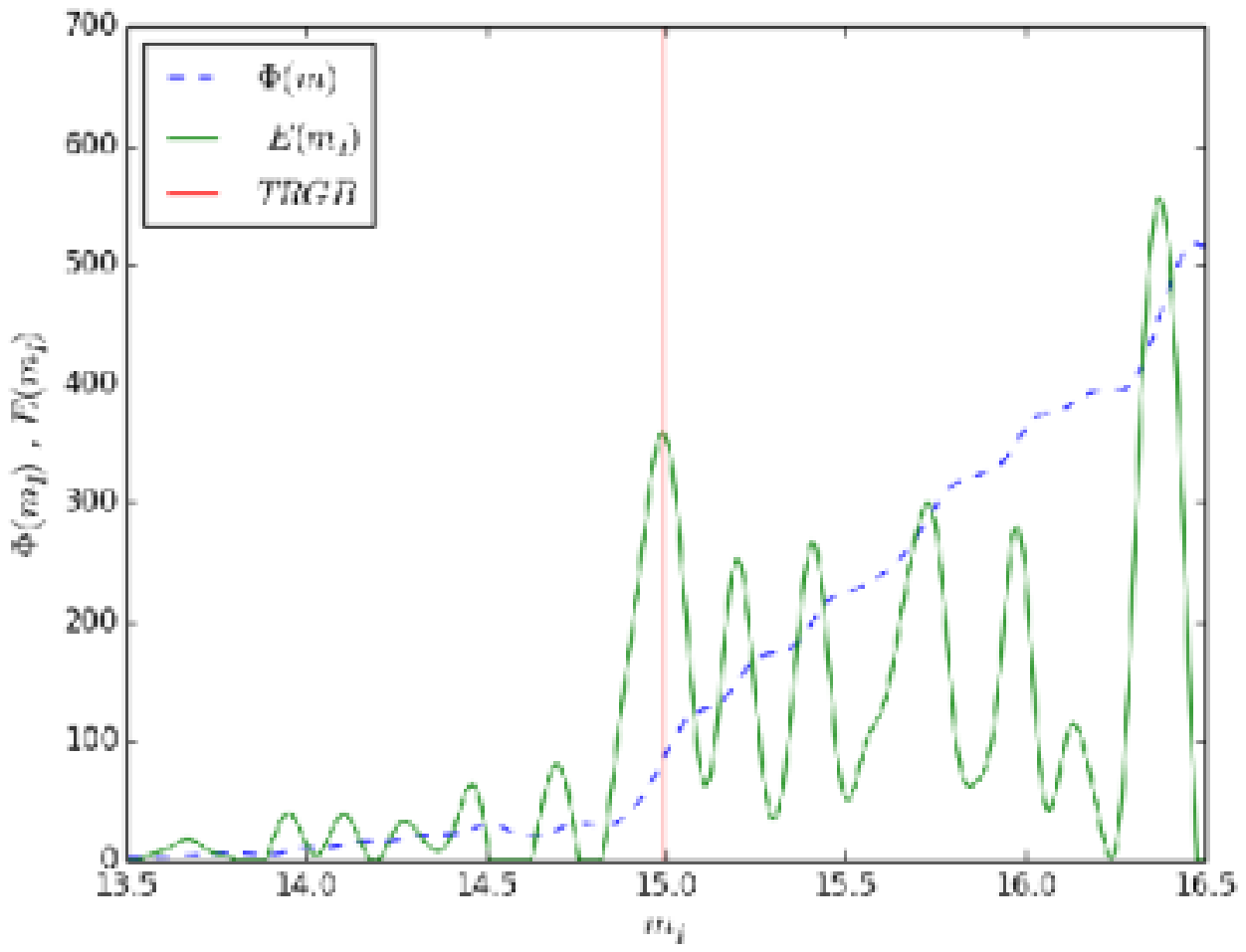}
\plotone{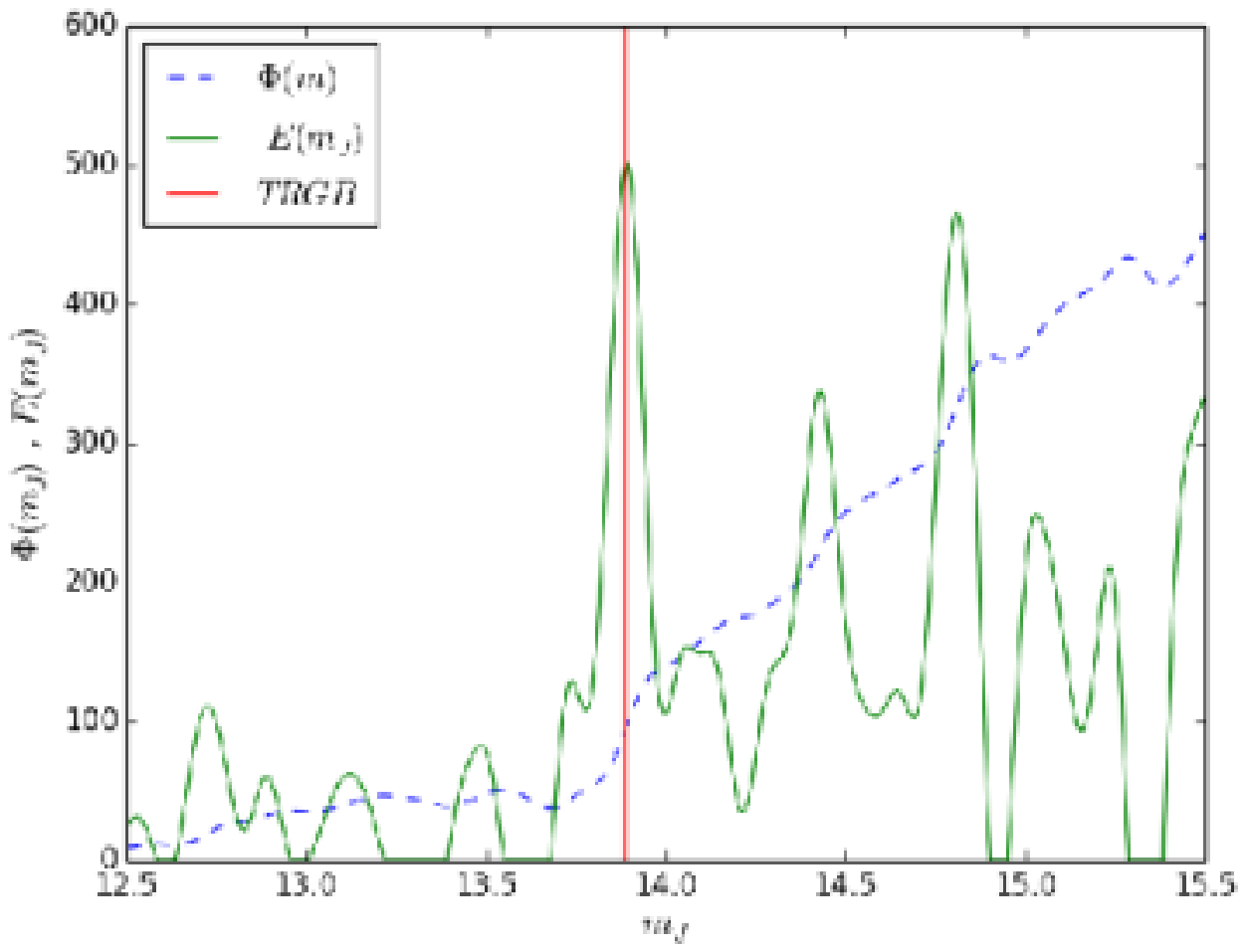}
\plotone{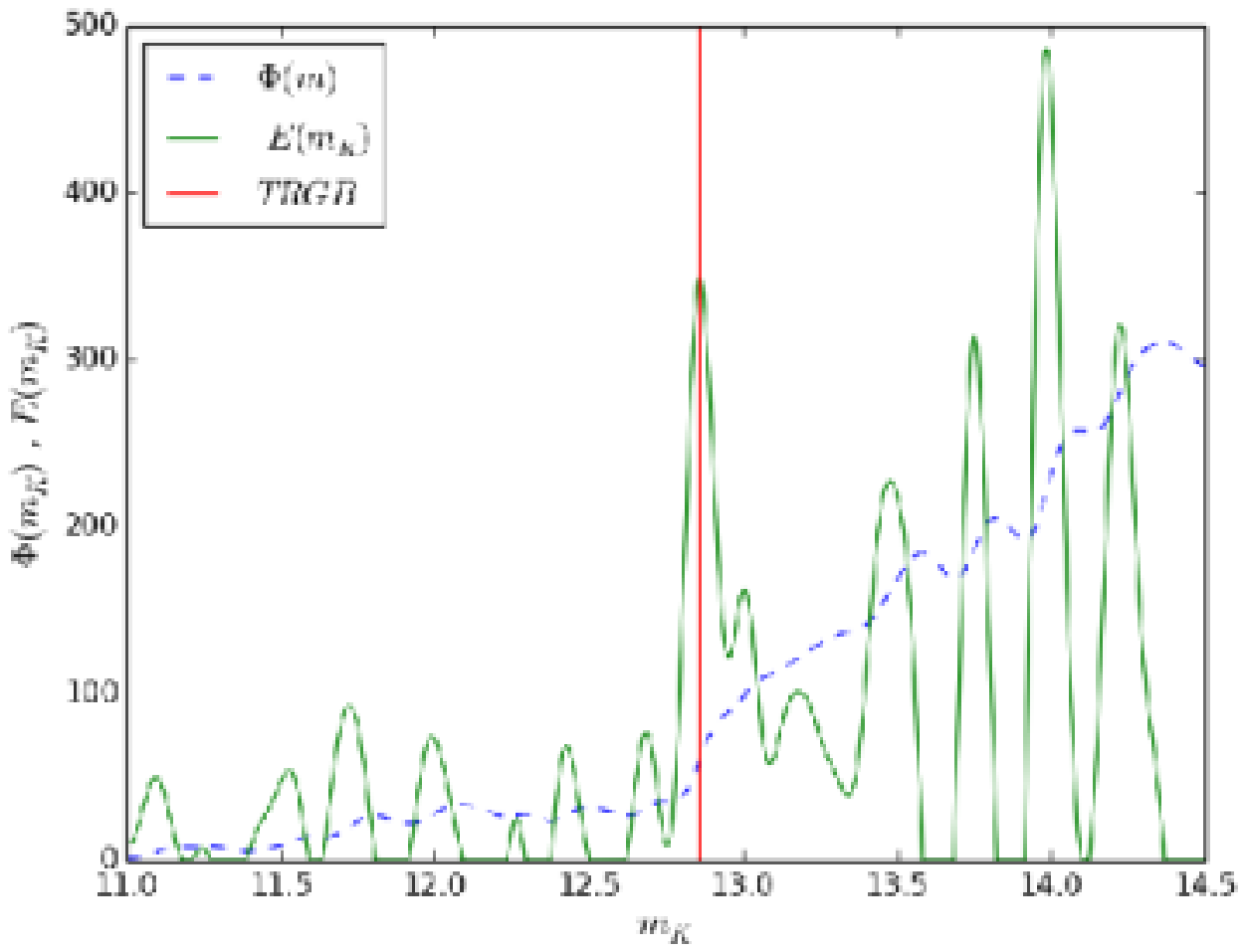}
\caption{The I, J and K band Gaussian-smoothed luminosity functions (blue dashed line) of the red giant branch in the Small Magellanic Cloud field smc106, 
and the corresponding outputs of the Sobel edge-detection filter (green solid line). The vertical red line indicates the detected TRGB magnitude.}
\label{fig:smc162}
\end{figure}

Other detection techniques offer some advantages, in particular when the RGB luminosity function is weakly populated (Makarov et al. 2006). These advantages are unessential for our data, because in each analyzed field at least 700 stars populate the range of 1 mag below the TRGB. Furthermore, one cannot exclude offsets between different methods. In order to estimate the statistical uncertainty of detection we set up Sobel filter with a Bootstrap resampling method. Figures \ref{fig:lmc106}, \ref{fig:lmc127} and \ref{fig:smc162} show exemplary luminosity functions and answers of the Sobel filter with marked TRGB detection for the LMC and SMC. Tables \ref{tab:lmc} and \ref{tab:smc} present the TRGB brightness measurements in the I, J and K bands in each of the 17 LMC fields and five SMC fields.

The mean magnitudes of the measured TRGB in the LMC are 14.62 $\pm$ 0.03 mag, 13.27 $\pm$ 0.04 mag and 12.13 $\pm$ 0.04 respectively in  the I, J and K bands. The mean magnitudes of the TRGB we measured in the SMC are 15.04 $\pm$ 0.07, 13.89 $\pm$ 0.04 and 12.91 $\pm$ 0.04 in the I, J and K bands.


\section{Reddening}
In order to accurately determine the distance one has to take into account the reddening. Extinction has a relatively small effect in the K infrared band, while it is crucial for the optical I band. 
Schlegel et al. (1998) give values of the Galactic foreground reddening toward the LMC and SMC: E(B-V)=0.057 mag and E(B-V)=0.035 mag, respectively. In case of the TRGB measurements, internal reddening is frequently neglected, due to the lack of knowledge of precise values. For the LMC and SMC, values of internal reddening are known from the color excess of red clump stars (Haschke, Grebel \& Duffau 2011). Like stars arriving toward the TRGB, red clump stars belong to Population II and it seems to be appropriate to adopt those values. For our investigation we decided to take into account only fields with internal reddening lower than E(V-I)=0.1 mag. The adopted values of E(V-I) from Haschke, Grebel \& Duffau (2011) are reported in Tables \ref{tab:lmc} and \ref{tab:smc}. Based on the Cardelli et al. (1989) $R_V = 3.1$ reddening law, we calculated values of extinction for our near-infrared and optical bands. The mean values for all our fields are: $A_V = 0.40$ mag,  $A_I = 0.24$ mag, $A_J = 0.11$ mag, $A_K = 0.05$ mag for the LMC, and $A_V = 0.20$ mag,  $A_I = 0.12$ mag, $A_J = 0.06$ mag, $A_K = 0.02$ mag for the SMC. 

\section{TRGB absolute brightness}
For the near-infrared bands we use the empirical TRGB absolute brightness-metallicity calibrations provided by
Valenti, Ferraro \& Origlia (2004). These calibrations were obtained from the homogeneous photometry of 24 Galactic globular clusters spanning a metallicity range from -2.12 to -0.49 dex. Absolute magnitudes are expressed in the 2MASS photometric system:

\begin{equation}
\label{eq:trgbJ}
M^\mathrm{TRGB}_\mathrm{J} =  -0.31 \mathrm{[Fe/H]} - 5.67 \mathrm ,
\end{equation}
\begin{equation}
\label{eq:trgbK}
M^\mathrm{TRGB}_\mathrm{K} =  -0.58 \mathrm{[Fe/H]} - 6.98 \mathrm .
\end{equation}

Valenti, Ferraro \& Origlia (2004) also provided a calibration for the absolute bolometric brightness of the TRGB in terms of metallicity:

\begin{equation}
\label{eq:trgbbol}
M^\mathrm{TRGB}_\mathrm{bol} = -0.18 \mathrm{[Fe/H]} - 3.87\mathrm .
\end{equation}

To obtain this calibration, the authors transformed the observed K band magnitudes onto a theoretical plane using 
bolometric corrections for Population II giants (Montegriffo et al. 1995). In the case of our data, we decided to transform the observed TRGB I band magnitudes to bolometric magnitudes with the bolometric corrections calculated from the unreddened (V-I) colors of the red giants (Da Costa \& Armandroff 1990):

\begin{equation}
\label{eq:bcI}
BC_I = 0.881-0.243(V-I)_0\mathrm.
\end{equation}

All of the above calibrations use metallicity on the Carretta \& Gratton (1997) scale. In the subsequent part of this paper we are also using metallicities calculated from the (V-I) color of the red giant branch that is expressed on the Zinn \& West (1984) scale. For this reason, in this case we employ the TRGB absolute bolometric brightness calibration of Bellazzini \& Ferraro (2001) in which metallicity is also on the Zinn \& West (1984) scale, and covers the range from -2.2 to -0.2 dex: 

\begin{equation}
\label{eq:trgbbol}
M^\mathrm{TRGB}_\mathrm{bol} = -0.12 \mathrm{[Fe/H]} - 3.76\mathrm .
\end{equation}

Bellazzini \& Ferraro (2001) also provided an optical I band TRGB calibration, that follows the formula:

\begin{equation}
\label{eq:trgbI}
M^\mathrm{TRGB}_\mathrm{I} = 0.14 \mathrm{[Fe/H]}^2 + 0.48 \mathrm{[Fe/H]} - 3.66 \mathrm{.}
\end{equation}

It is worth noting that the bolometric brightness is much less sensitive to metallicity than the K band brightness, but both brightnesses tend to increase with rising metallicity. The optical I band brightness follows the opposite trend.

\subsection{Metallicity determination}
The metallicity of the red giant branches in the LMC and SMC were the subject of many spectroscopic studies. 
Olszewski et al. (1991) and Da Costa et al. (1991) conducted measurements of the LMC clusters in the outer disc of the galaxy. They found old clusters with metallicity as low as -1.9 dex, while most of the examined objects had ages less than 3 Gyr, and metallicities in the vicinity of -0.5 dex. Surprisingly, there is an evident lack of clusters with intermediate age between 3 and 13 Gyr, and a corresponding metallicity from -1.7 to -1.0 dex.  
Cole at al. (2000) studied 39 red giants in the proximity of the LMC bar. In contradiction to Da Costa et al. (1991) and Olszewski et al. (1991), they found significantly fewer objects with metallicities lower than -1.0 dex, which hints at different chemical evolution scenarios in the central part of the LMC and its outer disc. Though the younger and more metal abundant part of the metallicity distribution seems to be similar for the Da Costa et al. (1991) measurements and the Cole at al. (2000) results. They both tend to have a metallicity maximum at -0.57 $\pm$ 0.04 dex and a mean metallicity of -0.64 $\pm$ 0.02 dex. On this basis, we adopt a metallicity of -0.6 dex with 0.1 dex error due to the standard deviation and the uncertainty of the calibration of the Ca II IR triplet for young, metal-rich stars. 

Since spectroscopic studies are time consuming, and for many galaxies are scarce, the common approach for the TRGB distance determinations is to adopt the metallicity calculated from the photometric calibration of the unreddened (V-I) color of the red giant branch (Lee at al. 1993):

\begin{equation}
\mathrm{[Fe/H]} =-12.64+12.6(V-I)_{-3.5}-3.3(V-I)^2_{-3.5} \mathrm{.} 
\label{eq:mett}
\end{equation}

The $(V-I)_{-3.5}$ color is measured for the absolute brightness range  $-3.4 > M_I > -3.6$ to assure sufficient number of stars (at least 100).  In order to determine these values we assumed 18.49 mag distance to the LMC (Pietrzy{\'n}ski et al. 2013) and 18.87 mag distance to the SMC (Graczyk et al. 2014). Reddening correction was applied to the I band brightness and (V-I) color of all the stars on the red giant branch. Measured mean $(V-I)_{-3.5}$ colors and their spread for all our fields are reported in Tables \ref{tab:lmc} and \ref{tab:smc}.

As pointed out by Salaris and Girardi (2005), in the case of the LMC, the photometric calibration leads to a much lower metallicity than the spectroscopic measurements. For our LMC fields, the mean metallicity calculated from the (V-I) color is -1.23 $\pm$ 0.07 dex. A similar value was also reported by Sakai et al. (2000). The cause of this discrepancy is related to the age distribution of the LMC stars and a different metal mixture than in globular clusters from which the above calibration was obtained (Salaris and Girardi 2005).

The same discrepancy in metallicity is observed in the SMC. Dobbie et al. (2014) measured spectroscopic metallicities for 3000 red giants in the SMC, and obtained a mean metallicity value of [Fe/H] = -1.0 $\pm$ 0.1 dex. Kamath et al. (2014) obtained a similar value ([Fe/H] = -1.14 $\pm$ 0.2 dex) from spectral observations of 42 post-RGB stars. Our calculations, based on photometric calibrations, lead to a metallicity -1.51 $\pm$ 0.03 dex. As in the case of the LMC, the metallicity calculated from photometric calibrations is lower by about 0.5 dex compared with the spectroscopic values. 

The calibrations introduced in the previous section are based on spectroscopic metallicities of stars in the Galactic globular clusters. For this reason we are using spectroscopic metallicity, while metallicity calibrated from the $(V-I)$ color will be discussed in the final part of this paper.


\section{Results}
Our measured mean TRGB brightnesses, adopted reddenings and metallicities lead to the following distance moduli for the LMC:

$$(m-M)_\mathrm{0,bol} = 18.66 \pm 0.03~\mathrm{(stat.)}~\pm 0.07~\mathrm{(syst.)}~\mathrm {mag}$$
$$(m-M)_\mathrm{0,I} = 18.29 \pm 0.03~\mathrm{(stat.)}~ \pm 0.08~\mathrm{(syst.)}~ \mathrm {mag}$$
$$(m-M)_\mathrm{0,J} = 18.63 \pm 0.04~\mathrm{(stat.)}~ \pm 0.05~\mathrm{(syst.)}~ \mathrm {mag}$$
$$(m-M)_\mathrm{0,K} = 18.70 \pm 0.04~\mathrm{(stat.)}~ \pm 0.07~\mathrm{(syst.)}~ \mathrm {mag}$$

For the SMC we obtain:

$$(m-M)_\mathrm{0,bol} = 19.14 \pm 0.07~\mathrm{(stat.)} \pm 0.07~\mathrm{(syst.)}~\mathrm {mag}$$
$$(m-M)_\mathrm{0,I} = 18.94 \pm 0.07~\mathrm{(stat.)} \pm 0.08~\mathrm{(syst.)}~\mathrm {mag}$$
$$(m-M)_\mathrm{0,J} = 19.16 \pm 0.04~\mathrm{(stat.)} \pm 0.05~\mathrm{(syst.)}~\mathrm {mag}$$
$$(m-M)_\mathrm{0,K} = 19.23 \pm 0.04~\mathrm{(stat.)} \pm0.07~\mathrm{(syst.)}~\mathrm {mag}$$

Tables \ref{tab:lmc_err} and \ref{tab:smc_err} summarize the contributions to total statistical and systematic uncertainties of our distance determinations. The systematic uncertainty of our distance determinations is composed of a reddening systematic error (0.05 mag) and a metallicity systematic error (0.1 dex). 
The statistical error is estimated as the standard deviation of the TRGB measurements for all fields. Identified contributors are photometric and detection errors, reddening errors, and variations of metallicity (0.08 dex for the LMC and 0.03 for the SMC). Metallicity variations were estimated from the mean color of the RGB with Equation 8. It is important to state at this moment, that all mentioned errors are not independent. Both photometric and reddening errors are correlated with estimated metallicity variations. 

Reddening is the main contributor to the statistical and systematic uncertainty of our distance determination for the I band and for the bolometric brightness, while it is negligible for the K band distance determination. The K band distance is mainly affected by the metallicity error, which is negligible for the bolometric brightness-based TRGB distance estimation.

\section{Discussion}
Our TRGB brightness measurements are in good agreement with results obtained by other authors, though some comment is necessary. Nikolaev \& Weinberg (2000) measured the TRGB brightness in the K band from 2MASS data, and obtained $K_{TRGB}=12.3 \pm 0.1$ mag. This value was estimated from the luminosity function histogram, which is biased by binning confines. In fact, direct examination of the presented histogram suggests that the value measured with Sobel filter should yield $K_{TRGB}=12.1$ mag and would be consistent with our result ($K_{TRGB} =12.13$ mag), and also with the recent result obtained by Macri et al. (2015) ($K_{TRGB}=12.11$ mag).
Cioni et al. (2000) measured the TRGB brightness from the DENIS survey, and obtained $K_{TRGB}=11.98 \pm 0.04$ mag in the DENIS photometric system. This estimation is in significant disagreement with Nikolaev \& Weinberg (2000), Macri et al. (2015) and our results. The reason for this disagreement, as indicated by Cioni et al. (2000), can be related to the non-standard detection technique used by the authors or, even more explicitly, to the non-standard photometric system of the DENIS survey. The I band TRGB brightness was obtained by Reid et al. (1987) ($I_{trgb} = 14.6$ mag) and Sakai et al. (2000) ($I_{trgb} = 14.54 \pm 0.04$ mag with reddening corrected individually for each star, with mean $A_I = 0.1$ mag).  Our I band TRGB brightness measurement ($I_{TRGB} = 14.62$ mag) is in a good agreement with these results. Nonetheless, reported distance moduli vary from 18.4 mag to 18.6 mag, due to different values of reddening and metallicity adopted by each author.

The distance determinations presented in our paper from the measurements of the TRGB brightness are inconsistent with the distances to the LMC and SMC obtained from the eclipsing binaries (LMC: 18.49 $\pm$ 0.02 mag, Pietrzy{\'n}ski et al. 2013; SMC: 18.97 $\pm$ 0.03 mag, Graczyk et al. 2014). Our optical I band distance to the LMC is underestimated at the level of 0.2 mag. The infrared K band distance is overestimated by 0.2 mag, both for the LMC and SMC. The discrepancy between I and K band distance moduli is 0.4 mag for the LMC and 0.3 mag for the SMC.

Although there are significant deviations for all bands, the differential distance ($\mu$) between the LMC and SMC is surprisingly consistent for the J, K bands and for the bolometric brightness, i.e. $\mu_J =0.53$ mag, $\mu_K =0.53$ mag and $\mu_{bol} = 0.48$ mag.It is also important that these values are in good agreement with the differential distance obtained from the eclipsing binaries ($\mu =0.48$ mag; Graczyk et al. 2014).  However, the optical I band differential distance between the LMC and SMC ( $\mu_{I} = 0.65$ mag) is not consistent with the J, K bands, bolometric brightness, and the eclipsing binaries.

Although a deep analysis of the observed discrepancies is beyond the main scope of this paper, we briefly discuss possible sources of systematic errors. One of the origins of systematic error may arise from the calibration of Valenti, Ferraro \& Origlia (2004). It affects distance estimations from infrared $J$ and $K$ band TRGB brightness, and from the bolometric brightness. In the LMC and SMC we expect that the Sobel filter method should indicate the TRGB brightness with a precision within the reported uncertainties, thanks to the well populated bright part of the RGB. In contrast to this, the calibrations of Valenti, Ferraro \& Origlia (2004) were obtained with {\bf the TRGB} brightness measurements in 24 Galactic globular clusters. For the majority of these globular clusters it is impossible to apply the Sobel filter, because in the close proximity of the TRGB there are only a few stars. Instead, the authors assumed that the brightest star on the RGB corresponds to the TRGB. This can lead to a significant systematic error of the absolute brightness of the TRGB in their calibration, which leads to an underestimation of our distance moduli based on the detection of the TRGB with the Sobel filter. Viaux et al. 2013 estimated the probability distribution of the offset between the brightest star in the M5 cluster and the TRGB. In the case of the M5 cluster, with a 65\% confidence level, the offset between the brightest star and the TRGB is smaller than 0.05 mag. To correctly asses this effect, a calculation for all clusters from Valenti, Ferraro \& Origlia (2004) should be applied, but it should not be larger than 0.05 mag, as in the case of the M5 cluster. We note at this point, that taking into account the offset between the brightest star on the RGB and the brightness of the TRGB will lead to even larger discrepancies of our distance measurements with the 18.49 mag and 18.97 mag value of the distance moduli to the LMC and SMC coming from the eclipsing binaries.

Another possible explanation of the disagreement of our results with the classical distances to the Magellanic Clouds is connected with the value of the zero-point in both calibrations (Valenti, Ferraro \& Origlia 2004 and Bellazzini \& Ferraro 2001). The error in the zero-point value may be caused by adopting incorrect values of the distances to the globular clusters which were used to obtain those calibrations. It would explain why the differential distance between the LMC and SMC is consistent in the J and K bands and for the bolometric brightness with the results from eclipsing binaries, but it does not explain why the distance obtained for each of these galaxies differs in the J and K bands and for the bolometric brightness. Even taking into account that the calibration of Valenti, Ferraro \& Origlia (2004) has a different error in the zero-point than the calibration of Bellazzini \& Ferraro (2001), it is unlikely for it to reach 0.4 mag difference between I and K bands.

While rather unlikely, one could suspect that the disagreement between bolometric brightness and K band is caused by adopting incorrect values of metallicity or reddening in our calculations.  A different value of metallicity (e.g. [Fe/H] = -1.0 dex) can bring the calculated distance to the LMC in the K band close to 18.50, but the discrepancies for the other bands remain at the level of 0.2 mag. Even adopting different values for both metallicity and reddening cannot bring all estimations together (I, J, K band and bolometric brightness) to the expected distance. The same applies to the SMC. This implies systematic errors, strongly depending on the band which is used for the measurements.

The most convincing explanation of the dominating systematic error is provided by Salaris and Girardi (2005). Their analysis of artificial synthesis populations based on star formation histories confirms the discrepancies obtained in this paper.
The authors are unable to predict precise corrections, but they show that young stars should produce an effect that is imitating a lower metallicity of the stars. The observed underestimation of the distance in the optical I band and overestimation in the near-infrared bands exactly corresponds to this effect.

As mentioned in the previous sections, the TRGB distance determinations are in many cases based on metallicities calculated from the (V-I) color of the red giant branch (equation 8). The metallicity calculated with this method yields values of -1.23 dex for the LMC, and -1.51 dex for the SMC, being 0.5 dex underestimated when comparing to the spectroscopic results.
The absolute brightness obtained from adopting this metallicity leads to distance moduli of 18.51 mag for the LMC, and 19.04 for SMC, and is in good agreement with the distance to the LMC and SMC obtained from the eclipsing binaries. This rather peculiar accuracy of distance estimation is caused by the opposing effects of stellar population to both calculated metallicity and bolometric brightness of the TRGB, which tend to compensate each other.  This effect will be discussed in detail in a forthcoming paper.

\section{Summary and Conclusions}
We have measured the brightnesses of the TRGB in the Large and Small Magellanic Clouds in the optical $I$ band and in the near-infrared $J$ and $K$ bands. Measurements were conducted separately in 17 fields in the central region of the LMC and in five fields in the center of the SMC. Variations of the TRGB brightness between different fields in the LMC are at the level of 0.03-0.04 mag and are caused by differences in the internal reddening and metallicity.

With the calibrations presented in Section 4, we were able to calculate absolute brightnesses of the TRGB in the $I$, $J$, $K$ bands, and for the bolometric brightness. The distances calculated on this basis are inconsistent between each other, and differ from the generally accepted 18.50 mag distance to the LMC, and 19.0 mag to the SMC. For the $I$ band, the distance is underestimated by 0.2 mag, and for the K band it is overestimated by 0.2 mag. The detected systematic errors are caused by population effects in the Large and Small Magellanic Clouds -- mainly by the mean age of the stars on the Red Giant Branch.

The reported differences can be taken as an upper limit on the systematic error of the TRGB distance measurements. They should not affect distance measurements to galaxies with Red Giant Branches dominated by a homogeneous and old stellar population.

This systematic population effect is smaller for the $J$ band and for the bolometric brightnesses, as compared to the K band. Moreover, distances obtained from the bolometric brightness in tandem with a metallicity calculated from the $(V-I)$ color of red giants agree with the 18.50 mag eclipsing binary distance to the LMC, and with the 19.00 mag distance to SMC. However, the adopted value of the metallicity is incorrect, and this fortuitous agreement should be considered as due to the compensatory effect of oppositely directed systematic errors of bolometric absolute brightness and metallicity.

\acknowledgments
We gratefully acknowledge financial support for this work from the Polish National Science Center grant PRELUDIUM 2012/05/N/ST9/03846 and TEAM subsidy from the Foundation for Polish Science (FNP). We also gratefully acknowledge financial support from the BASAL Centro de Astrofisica y Tecnologias Afines (CATA) PFB-06/2007, and from the Millenium Institute of
Astrophysics (MAS) of the Iniciativa Cientifica Milenio del Ministerio de Economia, Fomento y Turismo de Chile, project IC120009. M.C. acknowledges additional support from FONDECYT grant 1141141. This work is based on observations made by the OGLE project and the IRSF Magellanic Clouds Point Source Catalog, and it is a pleasure
to thank for the excellent photometric data provided by those teams.


\begin{figure}[htdp]
\plotone{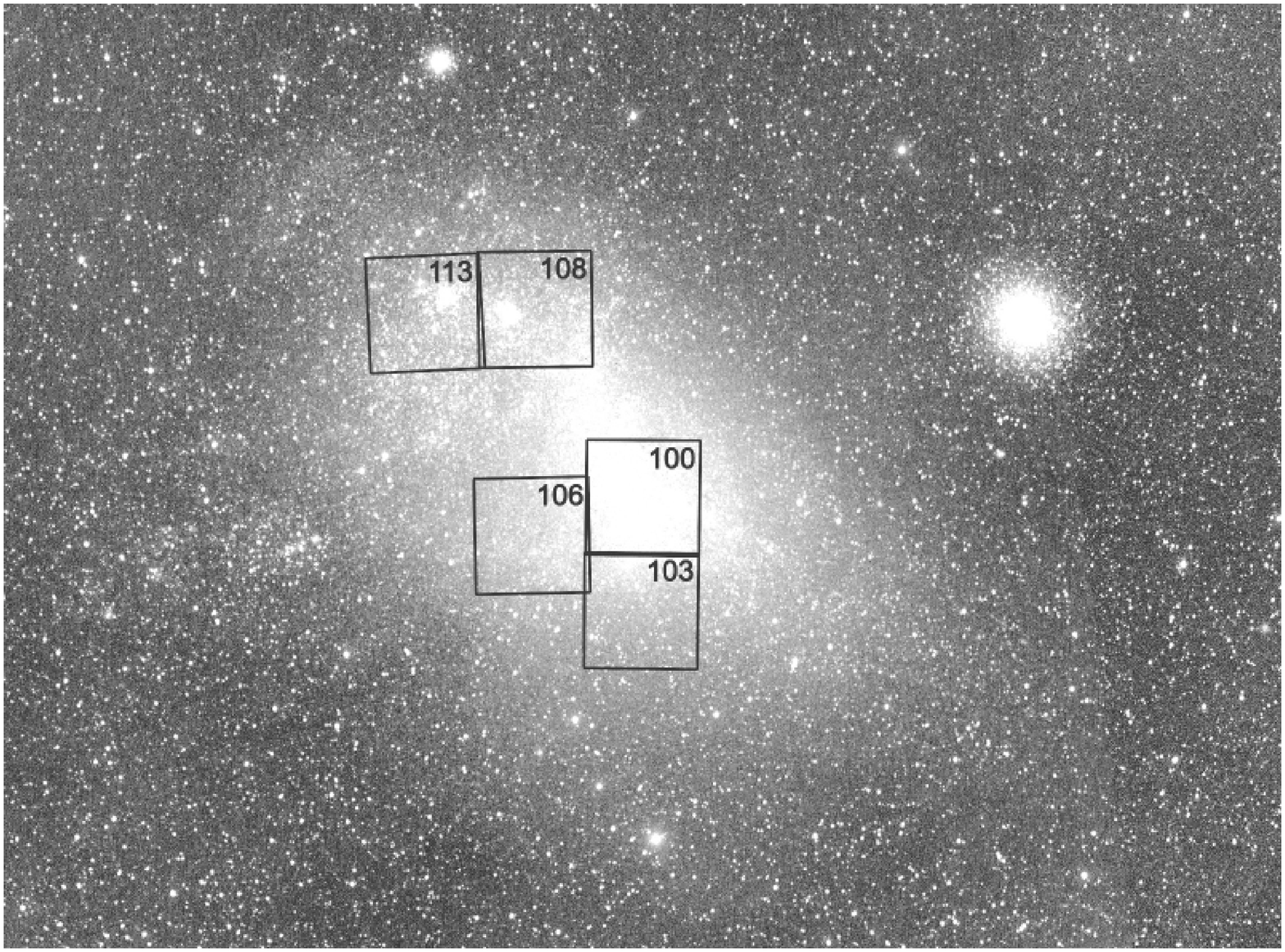}
\caption{The location of the 5 analyzed fields in the Small Magellanic Cloud. North is up, east is to the
left. Fields were selected to ensure a sufficiently large number of stars within 1 mag below the observed TRGB, and a minimal contribution to systematic errors from reddening and geometrical effects. The size of each field is 35 arcmin $\times$ 35 arcmin.}
\label{fig:smc}
\end{figure}


\begin{deluxetable}{ccccccccc}
\tablewidth{0pc}
\tablecaption{Summary information on the 17 analyzed fields in the Large Magellanic Cloud. 
For each field its center coordinates, the TRGB brightness in I, J, and K bands, the reddening E(V-I), and the averaged (V-I)$_{-3.5}$ color 
with its spread $\sigma_{(V-I)}$ are given.}

\tablehead{
\colhead{Filed} & \colhead{RA} & \colhead{DEC} & \colhead{I$_{TRGB}$} & \colhead{J$_{TRGB}$} & \colhead{K$_{TRGB}$} & \colhead{E(V-I)}  & \colhead{(V-I)$_{-3.5}$} &  \colhead{$\sigma_{(V-I)}$}   
}
\startdata
LMC100 & 5:19:02.2  & -69:15:07 & 14.61 & 13.22 & 12.12 & 0.06 & 1.71 & 0.18 \\
LMC103 & 5:19:02.9 & -69:50:26 & 14.62 & 13.24 & 12.11 & 0.06 & 1.69 & 0.18\\
LMC108 & 5:13:01.9 & -67:28:40 & 14.57 & 13.31 & 12.14 & 0.09 & 1.68 & 0.17\\
LMC110 & 5:12:43.6 & -68:39:42 & 14.58 & 13.25 & 12.11 & 0.08 & 1.70 & 0.18\\
LMC112 & 5:12:21.5 & -69:50:21 & 14.59 & 13.27 & 12.10 & 0.07 & 1.69 &  0.14\\
LMC117 & 5:06:55.3 & -68:03:58 & 14.64 & 13.27 & 12.17 & 0.05 & 1.63 & 0.13\\
LMC118 & 5:06:25.4 & -68:39:25 & 14.61 & 13.27 & 12.14 & 0.07 & 1.65 & 0.15\\
LMC119 & 5:06:02.5 & -69:15:02 & 14.65 & 13.27 & 12.13 & 0.07 & 1.63 & 0.15 \\
LMC120 & 5:05:39.8 & -69:50:28 & 14.65 & 13.35 & 12.15 & 0.08 & 1.67 & 0.12 \\
LMC126 & 5:00:02.4 & -68:39:31 & 14.61 & 13.29 & 12.18 & 0.08 & 1.65 & 0.16 \\
LMC127 & 4:59:33.6 & -69:14:54 & 14.63 & 13.29 & 12.21 & 0.09 & 1.66 & 0.15 \\
LMC128 & 4:59:03.6 & -69:50:24 & 14.68 & 13.25 & 12.11 & 0.08 & 1.67 & 0.12 \\
LMC161 & 5:25:32.5 & -69:14:59 & 14.62 & 13.34 & 12.17 & 0.09 & 1.67 & 0.15 \\
LMC162 & 5:25:43.3 & -69:50:24 & 14.57 & 13.22 & 12.08 & 0.05 & 1.65 & 0.16 \\
LMC163 & 5:25:52.2 & -70:25:55 & 14.64 & 13.23 & 12.07 & 0.07 & 1.73 & 0.12\\
LMC169 & 5:32:22.8 & -69:50:26 & 14.69 & 13.30 & 12.15 & 0.08 & 1.72 & 0.16\\
LMC170 & 5:32:48.1 & -70:25:53 & 14.60 & 13.24 & 12.10 & 0.06 & 1.66 & 0.18\\
\enddata
\label{tab:lmc}
\end{deluxetable}


\begin{deluxetable}{ccccccccc}
\tablewidth{0pc}
\tablecaption{Summary information on the 5 analyzed fields in the Small Magellanic Cloud. 
For each field its center coordinates, the TRGB brightness in I, J, and K bands, the reddening E(V-I), and the averaged color (V-I)$_{-3.5}$ color 
with it's spread $\sigma_{(V-I)}$ are given.}

\tablehead{
\colhead{Filed} & \colhead{RA} & \colhead{DEC} & \colhead{I$_{TRGB}$} & \colhead{J$_{TRGB}$} & \colhead{K$_{TRGB}$} & \colhead{E(V-I)}  & \colhead{(V-I)$_{-3.5}$} &  \colhead{$\sigma_{(V-I)}$}   
}

\startdata
SMC100 & 0:50:06.4 & -73:08:19 & 15.20 & 13.87 & 12.75 & 0.04 & 1.50 & 0.11\\
SMC103 & 0:50:08.6 & -73:43:44 & 15.15 & 13.84 & 12.82 & 0.04 & 1.50 & 0.11 \\
SMC106 & 0:58:06.7 & -73:20:21 & 14.99 & 13.89 & 12.86 & 0.04 & 1.48 & 0.07\\
SMC108 & 0:57:31.5 & -72:09:29 & 15.02 & 13.91 & 12.98 & 0.04 & 1.48 & 0.09\\
SMC113 & 1:05:02.8 & -72:09:32 & 14.98 & 13.92 & 12.90 & 0.04 & 1.49 & 0.11 \\
\enddata
\label{tab:smc}
\end{deluxetable}


\begin{deluxetable}{ccccc}
\tablewidth{0pc}
\tablecaption{Contributions to the statistical and systematic uncertainties on the TRGB distance measurements to the Large Magellanic Cloud in I, J, K bands and for the bolometric brightness.
}

\tablehead{
\colhead{} & 
\colhead{Bol} & 
\colhead{I} & 
\colhead{J} & 
\colhead{K}  
}

\startdata
 \multicolumn{5}{c}{statistical erros}\\
 \hline
STD  & 0.03 & 0.03 & 0.04 & 0.04 \\
\hline
photometric & 0.02 & 0.02 & 0.02 & 0.02\\
detection & 0.02 & 0.02 & 0.02 & 0.02 \\
STD reddening & 0.02 & 0.01 & 0.006 & 0.001 \\
STD metallicity & 0.01 & 0.03 & 0.02 & 0.05 \\
\hline
\multicolumn{5}{c}{systematic errors}\\
\hline
reddening & 0.07 & 0.07 & 0.04 & 0.02 \\
metallicity & 0.02 & 0.04 & 0.03 & 0.06 \\ 
\hline

\enddata
\label{tab:lmc_err}
\end{deluxetable}


\begin{deluxetable}{ccccc}
\tablewidth{0pc}
\tablecaption{Contribution to the statistical and systematic uncertainties on the TRGB distance measurements to the Small Magellanic Cloud in I, J, K bands and for the bolometric brightness.}

\tablehead{
\colhead{} & 
\colhead{Bol} & 
\colhead{I} & 
\colhead{J} & 
\colhead{K}  
}

\startdata
 \multicolumn{5}{c}{statistical errors}\\
 \hline
STD  & 0.07 & 0.07 & 0.04 & 0.04 \\
\hline
photometric & 0.02 & 0.02 & 0.02 & 0.02\\
detection & 0.02 & 0.02 & 0.02 & 0.02 \\
STD reddening & 0.02 & 0.01 & 0.006 & 0.001 \\
STD metallicity & 0.005 & 0.01 & 0.01 & 0.02 \\
\hline
\multicolumn{5}{c}{systematic errors}\\
\hline
reddening & 0.07 & 0.07 & 0.04 & 0.02 \\
metallicity & 0.02 & 0.04 & 0.03 & 0.06 \\ 
\hline

\enddata
\label{tab:smc_err}
\end{deluxetable}

\end{document}